\documentclass[journal=jctcce,manuscript=article,layout=twocolumn]{achemso}
\pdfoutput=1 
\usepackage[T1]{fontenc}

\usepackage{graphicx}
\usepackage{color}
\usepackage[normalem]{ulem}
\usepackage{bm}

\usepackage{amsmath}
\usepackage{amssymb}

\title{Efficient analytic second derivative of electrostatic embedding QM/MM energy: normal mode analysis of plant cryptochrome}
\author{Karno Schwinn}
\affiliation{Aix-Marseille Univ, CNRS, ICR, Marseille, France}
\author{Nicolas Ferr{\'e}}
\affiliation{Aix-Marseille Univ, CNRS, ICR, Marseille, France}
\author{Miquel Huix-Rotllant}
\affiliation{Aix-Marseille Univ, CNRS, ICR, Marseille, France}
\email{miquel.huixrotllant@univ-amu.fr}

\let\oldmaketitle\maketitle
\let\maketitle\relax

\begin{document}

\twocolumn[
\begin{@twocolumnfalse}
\oldmaketitle
\begin{abstract}
Analytic second derivatives of electrostatic embedding (EE) quantum mechanics/molecular mechanics (QM/MM) energy are important for performing vibrational analysis and simulating vibrational spectra of quantum systems interacting with an environment represented as a classical electrostatic potential. The main bottleneck of EE-QM/MM second derivatives is the solution of coupled perturbed equations for each MM atom perturbation. Here, we exploit the Q-vector method {[J. Chem. Phys., 151, 041102 (2019)]} to workaround this bottleneck. We derive the full analytic second derivative of the EE-QM/MM energy, which allows to compute QM, MM and QM-MM Hessian blocks in an efficient and easy to implement manner. To show the capabilities of our method, we compute the normal modes for the full \textit{arabidopsis thaliana} plant cryptochrome. We show that the flavin adenine dinucleotide vibrations (QM subsystem) strongly mix with protein modes. We compute approximate vibronic couplings for the lowest bright transition, from which we extract spectral densities and the homogeneous broadening of FAD absorption spectrum in protein using vibrationally resolved electronic spectrum simulations.
\end{abstract}
\end{@twocolumnfalse}
]

\maketitle

\section{Introduction}

Analytic derivatives of the energy revolutionized quantum chemistry,\cite{Pulay14} giving access to gradients and Hessians (respectively first and second derivative with respect to nuclear positions) to explore potential energy surfaces,\cite{Pulay69} in addition to derivatives with respect to electromagnetic fields, from which properties are nowadays routinely extracted.

The development of mixed quantum mechanics (QM) and molecular mechanics (MM) models by Warshel, Levitt and Karplus, allowed a more realistic simulation of quantum systems embedded in complex environments.\cite{Warshel76,Field90,Senn_09,Nobel_13} Analytic derivatives of QM/MM energy have essentially focused on analytic gradient, which have been implemented for the main self-consistent field (SCF) and post-SCF QM methods.\cite{Kongsted2002,Sinicropi2008,Ferre02,Melaccio2011,Zhang2010,Tu2010,Barone2010,Carnimeo2015} Analytic second derivatives of QM/MM energy are less commonly available. As a matter of fact, most of the QM/MM Hessian constructions are based on the so-called Partial Hessian approach (see  \citenum{Mroginski_13} and references therein) in which only the QM degrees of freedom are considered. A first derivation of an analytic Hessian was introduced by Morokuma and coworkers in the framework of the substractive ONIOM QM/MM scheme\cite{Dapprich99}. One year later, Cui and Karplus proposed an analytic Hessian evaluation in their additive QM/MM approach.\cite{Cui00,Cui00b} The latter method is expensive, mainly due to the fact that the Hessian calculation involves solving a coupled perturbed (CP) SCF equation not only for the QM atoms but also for the MM atoms. Thus, the method of Cui and Karplus resorted to approximations of the CPSCF equations. More recently, Ghysels, Woodcock III and coworkers developed a coarse-graining method for the MM environment to reduce the number of CPSCF equations to solve.\cite{Ghysels11} A Hessian with fragment molecular orbital method has been derived by Nakata and coworkers.\cite{Nakata13} Other approximate second derivative methods have been proposed by Sakai and Morita,\cite{Sakai05} and Gadre and coworkers.\cite{Rahalkar08} The most up-to-date formulation of the Hessian is certainly the one by Capelli and coworkers, especially derived for their very promising polarizable QM/MM method which can be used to calculate infra-red spectra of very large systems in the so-called Partial Hessian Vibrational Approach, hence ignoring the major part of the Hessian that only depends on the MM atoms.\cite{Giovannini_19a}

Herein, we develop a new method for calculating the complete second derivatives of the QM/MM energy with additive QM-MM interaction terms. We base our derivation on single-determinant SCF wavefunctions characterizing a QM charge density interacting with a classical point-like atomic environment via the Electrostatic Potential Fitting (ESPF) interaction Hamiltonian developed by one of us.\cite{Ferre02,Melaccio2011} For that purpose, we apply the CPSCF Q-vector method for MM atoms,\cite{Schwinn19} avoiding one of the main bottlenecks of second derivatives of QM-MM energies. The method works even when the QM and MM regions are covalently linked via the use of the link atom method of Morokuma and coworkers.\cite{Dapprich99} For testing our implementation, we apply this method to perform a normal mode analysis of \textit{arabidopsis thaliana} cryptochrome,\cite{Muller14} and we compute the vibronic couplings, which allows us to efficiently construct spectral densities and account of homogeneous broadening of the spectrum via vibrationally resolved spectrum simulations.

\section{Analytic Derivatives of the ESPF QM/MM energy}

We use the notation ${\bf R}={\left(x,y,z\right)}$ to symbolize the coordinates of QM atom, and ${\tilde{\bf R}}={\left({\tilde{x}},{\tilde{y}},{\tilde{z}}\right)}$ to symbolize the coordinates of a MM atom. The partial derivatives with respect to a coordinate will be written in the short-hand notation
\begin{equation}
E^x=\frac{\partial E}{\partial x} 
\end{equation}
The indexes $i,j,\cdots$ will refer to occupied molecular orbitals (MO), $a,b,\cdots$ will refer to virtual MOs, $p,q,\cdots$ to generic MOs (both occupied and virtual), and $\mu,\nu,\cdots$ to atomic orbital (AO) index. Capital roman letters will be used for QM atom indexes.

\subsection{ESPF QM/MM energy}

In the ESPF method, the QM/MM energy can be written as
\begin{equation}
\label{eq:totener}
E=E_{QM}+E_{MM}
\end{equation}
in which $E_{QM}$ is the internal energy of the QM subsystem whose electron density is polarized by the MM subsystem (thus containing the energy contribution due to the electrostatic interaction with the MM subsystem), while $E_{MM}$ is the internal energy of the MM part complemented with classical interaction terms between QM and MM atoms. The electrostatic interaction energy in the ESPF method is evaluated as $\Delta E=\mathrm{Tr}\left[{\bf P}{\bf h}\right]$, in which ${\bf P}$ is the atomic density and ${\bf h}$ is the QM/MM interaction hamiltonian.\cite{Ferre02,Melaccio2011} This interaction energy is also expressed as the classical electrostatic interaction between quantum-derived atomic multipoles $q_A$, $\mu_A$, $\dots$ and the MM electrostatic potential $\phi_A$, the electric field vector $\mathbf{\nabla}\phi_A$, $\dots$ 
\begin{eqnarray}
\label{eq:interacener}
\Delta E& = &\sum_{A} \left ( q_{A}\phi_{A} - {\boldsymbol\mu}_{A}\cdot\nabla\phi_{A} + \dots \right)  \nonumber \\
 & = & \Delta E^{0} + \Delta E^{1} + \dots 
\end{eqnarray}
in which the index $A$ runs over the number of QM atoms $N_{QM}$. For the sake of simplicity, we will restrict the multipolar expansion (Eq. \ref{eq:interacener}) to the order 0 in the following, even though our method can be easily generalized to higher orders.\cite{Schwinn19}  The expressions for the derivatives of the QM-only and MM-only energy terms will be omitted here, since they have been lengthy discussed in the literature.\cite{Pople1979,Ferre02,Kaledin2006} Here, we will focus on the electrostatic interaction term and its derivatives.

At the order 0 of the QM density of charges, each atomic charge $q_A$ is expressed as the sum of the nuclear charge $Z_A$ and the mean value of an atomic one-electron operator, whose matrix elements are denoted $Q_{A,\mu\nu}$. The interaction energy in atomic orbital basis reads,
\begin{eqnarray}
\label{eq:eespfinteract}
\Delta E&=&\sum_{A,\mu\nu}{\left(-P_{\mu\nu}Q_{A,\mu\nu}+Z_A\right)\phi_A} \\
&=&\sum_{A}{q_A\phi_A}=\sum_{\mu\nu}{P_{\mu\nu}h_{\mu\nu}}+\sum_A{Z_A\phi_A} \nonumber
\end{eqnarray}
in which we have introduced a new one-electron operator whose elements are:,
\begin{equation}
\label{eq:hespf}
h_{\mu\nu} =-\sum_A{Q_{A,\mu\nu} \phi_A}
\end{equation}
In a nutshell, the ESPF method consists in fitting the $Q_{A,\mu\nu}$ matrix elements to reproduce the purely electronic QM electrostatic integrals calculated on a grid surrounding the QM subsystem\cite{Ferre02}.

\subsection{First derivative}

The analytic first derivatives of the ESPF interaction term were first derived in Ref.\ \citenum{Ferre02}. For completeness, we reexpress them with the current notation. The first derivative of Eq.\ \ref{eq:eespfinteract} can be expressed as a vector gradient,
\begin{equation}
\nabla [\Delta E]=\begin{bmatrix}
\Delta E^x \\
\Delta E^{\tilde{x}}
\end{bmatrix}
\end{equation}
in which the first block corresponds to the derivative with respect to QM atom positions and the second block to MM atoms. The expression of the energy gradient with respect to QM coordinates is given by
\begin{equation}
\Delta E^x=\sum_{\mu\nu}{P_{\mu\nu}^{x}h_{\mu\nu}}+\sum_{\mu\nu}{P_{\mu\nu}h_{\mu\nu}^{x}} +\sum_A{Z_A\phi_A^x}
\end{equation}
As usual, the first term is not calculated thanks to the introduction of the so-called energy-weighted density matrix $-{\bf P}{\bf h}{\bf P}{\bf S}^x$.\cite{Pople_79,Ferre02,Melaccio2011}  The derivative of the ESPF operator element $h_{\mu\nu}$ is given in Appendix \ref{sec:app1}. Regarding the derivative with respect to the MM coordinates, the energy gradient can be expressed as
\begin{equation}
\Delta E^{\tilde{x}}=\sum_{A}{q_A\phi_A^{\tilde{x}}} \, ,
\end{equation}
noting that ${\bf S}^{\tilde{x}}=0$ for MM perturbations.

\subsection{Second derivative}

The analytic second derivative of Eq.\ \ref{eq:eespfinteract} can be expressed as a Hessian matrix
\begin{eqnarray}
&&\nabla^2[\Delta E]=
\begin{bmatrix}
\Delta E^{xy} & \Delta E^{x\tilde{y}} \\
\Delta E^{\tilde{x}y} & \Delta E^{\tilde{x}\tilde{y}}
\end{bmatrix}
\end{eqnarray}
in which the $\Delta E^{xy}$ is the block of second derivatives with respect to QM atom coordinates, $\Delta E^{\tilde{x}\tilde{y}}$ is the second derivative block with respect to MM atom coordinates, and $\Delta E^{x\tilde{y}}=\Delta E^{\tilde{x}y}$ is the second derivative block with respect to one MM and one QM atom coordinates.

The analytic formula for the $\Delta E^{xy}$ second derivative is given by
\begin{equation}
\label{eq:hessblock1}
\Delta E^{xy}=\sum_A{Z_A\phi_A^{xy}}+\sum_{\mu\nu}{\left(P_{\mu\nu}h^{xy}_{\mu\nu}+P^y_{\mu\nu}h_{\mu\nu}^{x}\right)}
\end{equation}
The second derivative of the Fock operator requires the second derivative of the ESPF operator, which is given in \ref{sec:app1}. This second derivative block also requires the density matrix derivative ${\bf P}^y$ with respect to QM atom coordinates, which is obtained by solving a CPSCF equation for each QM coordinate.\cite{McWeeny1960}

The $\Delta E^{x\tilde{y}}$ block over a QM and an MM atom coordinates is given by
\begin{equation}
\label{eq:hessblock2}
\Delta E^{x\tilde{y}}=\sum_A{Z_A\phi_A^{x\tilde{y}}}+\sum_{\mu\nu}{\left(P_{\mu\nu}h^{x\tilde{y}}_{\mu\nu}+P^{\tilde{y}}_{\mu\nu}h^{x}_{\mu\nu}\right)} 
\end{equation}
The explicit form of the second derivative of the ESPF operator is given in Appendix \ref{sec:app1}. This energy derivative requires the explicit construction of ${\bf P}^{\tilde{x}}$. These would require the solution of a CPSCF equation for all MM atom coordinates, which is prohibitive. A more convenient form to implement, is to compute the equivalent $\Delta E^{\tilde{x}y}$ derivative block, which is given by
\begin{equation}
\label{eq:hessblock2b}
\Delta E^{\tilde{x}y}= \sum_A{\left(q^{y}_A\phi_A^{\tilde{x}}+q_A\phi_A^{\tilde{x}y}\right)} 
\end{equation}
in which the charge derivative is given by
\begin{equation}
q^{y}_A=-\sum_{\mu\nu}{\left(P^y_{\mu\nu}Q_{A,\mu\nu}+P_{\mu\nu}Q^y_{A,\mu\nu}\right)} 
\end{equation}
In this case, only derivatives of the density matrix with respect to QM coordinates are required, which is already constructed for the QM energy second derivative.\cite{Pople1979}

Finally, the $\Delta E^{\tilde{x}\tilde{y}}$ derivatives over MM atom coordinates is given by
\begin{equation}
\label{eq:hessblock3}
\Delta E^{\tilde{x}\tilde{y}}=\sum_{A}{\left(q_A^{\tilde{y}}\phi^{\tilde{x}}_{A}+q_A\phi_A^{\tilde{x}\tilde{y}}\right)} 
\end{equation}
requires the derivatives of the ESPF atomic charges, which would in principle require the density derivatives 
\begin{equation}
\label{eq:qderiv}
q^{\tilde{y}}_A=-\sum_{\mu\nu}{P^{\tilde{y}}_{\mu\nu}Q_{A,\mu\nu}} 
\end{equation}

The construction of ${\bf P}^{\tilde{y}}$ would imply solving a coupled-perturbed type equation for each MM perturbation. Usually, the number of MM atoms is hundreds to thousands, becoming rapidly intractable. In Ref.\ \citenum{Schwinn19}, we showed how to avoid this by defining an auxiliary set of coupled-perturbed equations (Q-vector method) that only scales with the number of QM atoms, and from which we can efficiently reconstruct the charge derivatives,
\begin{eqnarray}
q^{\tilde{y}}_A=-\sum_{B,ia}{Q_{A,ia}}\phi_B^{\tilde{y}}{\tilde Q}_{B,ia} \, .
\end{eqnarray}
where ${\tilde Q}_{B,ia}$ is the solution of the Q-vector coupled-perturbed equation,\cite{Schwinn19} which allows to rewrite the MM contribution to second derivative as
\begin{eqnarray}
\Delta E^{\tilde{x}\tilde{y}}=\sum_{A}{\left(q_A\phi_A^{\tilde{x}\tilde{y}}+\sum_{B,ia}{Q_{A,ia}{\tilde{Q}}_{B,ia}\phi_B^{\tilde{y}}\phi^{\tilde{x}}_{A}}\right)}   
\end{eqnarray}
This equation can be interpreted in terms of classical atomistic mechanics,
\begin{eqnarray}
\Delta E^{\tilde{x}\tilde{y}}=\sum_{A}{\left(q_A\phi_A^{\tilde{x}\tilde{y}}+\sum_{B}{Q_{A}Q'_{AB}\phi_B^{\tilde{y}}\phi^{\tilde{x}}_{A}}\right)}   
\end{eqnarray}
in which $Q_A=-Tr[{\bf P}{\bf Q}_A]$, and
\begin{equation}
\label{eq:qtensor}
Q'_{AB}=\frac{\sum_{ia}{{\tilde{Q}}_{B,ia}Q_{A,ia}}}{Q_A}
\end{equation}
Then, $Q'_{AB}$ can be interpreted as a tensor that, given an electric charge $Q_A$ on QM atom A interacting with a gradient of the external potential, will give the ``effective'' charge on QM center B that interacts with another gradient of the external field.

\subsection{Link atoms}

For molecular systems in which there exist covalent bonds at the interface between the MM and QM subsystems, unphysical free valences are created in the QM side. Morokuma and coworkers proposed to use a fictitious link atom in order to saturate these dangling bonds, most often using a hydrogen atom. The link atom is constrained to remain positioned between the frontier MM and QM atoms,\cite{Dapprich99}
\begin{equation}
{\bf R}_{link}={\bf R}_{QM}+g\left({\bf {\tilde{R}}}_{MM}-{\bf R}_{QM}\right) 
\end{equation}
in which $g$ is a parameter that fixes the coordinates of the link atom between the covalently bonded QM and MM atoms. 

From the QM perspective, the ``auxiliary'' link atom is considered as a QM atom, and therefore, N$_{QM}$ includes all real QM atoms plus the link atoms. Consequently, there will be an extra contribution for the QM and MM frontier atoms that comes from the link atom. In the case of first energy derivatives, this is given by
\begin{eqnarray}
E^{x}|_{link}&=&E^x+(1-g)\frac{\partial E}{\partial x_{link}} \nonumber \\
E^{\tilde{x}}|_{link}&=&E^{\tilde{x}}+g\frac{\partial E}{\partial \tilde{x}_{link}}  
\end{eqnarray}
In other words, the link atom gradient is projected onto the frontier QM and MM atoms.
Similarly, the contribution of the link atom to the MM and QM second derivatives can be expressed as
\begin{eqnarray}
E^{xy}|_{link}&=&E^{xy}+(1-g)^2\frac{\partial^2 E}{\partial x_{link}\partial y_{link}} \nonumber \\
&+&(1-g)\left(\frac{\partial^2 E}{\partial x\partial y_{link}}+\frac{\partial^2 E}{\partial x_{link}\partial y}\right) \nonumber \\
E^{x\tilde{y}}|_{link}&=&E^{x\tilde{y}}+g(1-g)\frac{\partial^2 E}{\partial x_{link}\partial y_{link}} \nonumber \\
&+&(1-g)\frac{\partial^2 E}{\partial x_{link}\partial{\tilde{y}}}+g\frac{\partial^2 E}{\partial x\partial y_{link}} \nonumber \\
E^{\tilde{x}y}|_{link}&=&E^{\tilde{x}y}+g(1-g)\frac{\partial^2 E}{\partial x_{link}\partial y_{link}} \nonumber \\
&+&g\frac{\partial^2 E}{\partial x_{link}\partial y}+(1-g)\frac{\partial^2 E}{\partial \tilde{x}\partial y_{link}} \nonumber \\
E^{\tilde{x}\tilde{y}}|_{link}&=&E^{\tilde{x}\tilde{y}}+g^2\frac{\partial^2 E}{\partial x_{link}\partial y_{link}} \nonumber \\
&+&g\left(\frac{\partial^2 E}{\partial {\tilde{x}}\partial y_{link}}+\frac{\partial^2 E}{\partial x_{link}\partial {\tilde{y}}}\right)
\end{eqnarray}

\section{Computational Details}

All electronic structure computations have been performed with a modified version of {\sc Gaussian16} including the Q-vector method and the ESPF energy, gradient and Hessian. The MM information is obtained via an interface with a modified version of {\sc Tinker} 6.3.3.\cite{g16,tinker,Ferre02} 

The vibrationally resolved spectrum for FAD has been obtained by Fourier transforming the time-dependent auto-correlation function for the multi-dimensional harmonic oscillator, as implemented in {\sc Gaussian16}.\cite{Baiardi13,g16} The Franck-Condon approximation of the auto-correlation function has been integrated on 2$^{20}$ time steps, over an interval of about 10.5 ps. A Gaussian broadening of 235 cm$^{-1}$ has been applied to the vibrational peaks. 

All QM calculations have been performed at the DFT and full TDDFT levels,\cite{Casida2012} using the B3LYP functional\cite{B3,LYP,VWN} and the TZVP/TZVPFit basis set.\cite{Schaefer1992,Schaefer1994,Eichkorn1995,Eichkorn1997} For the MM calculations, we have employed the {\sc AMBER99} force field.\cite{amber18} Van der Waals parameters for FAD have been taken from Ref. \citenum{Antony00}.

\section{Results and Discussion}

\begin{figure}
    \includegraphics[width=\linewidth,keepaspectratio]{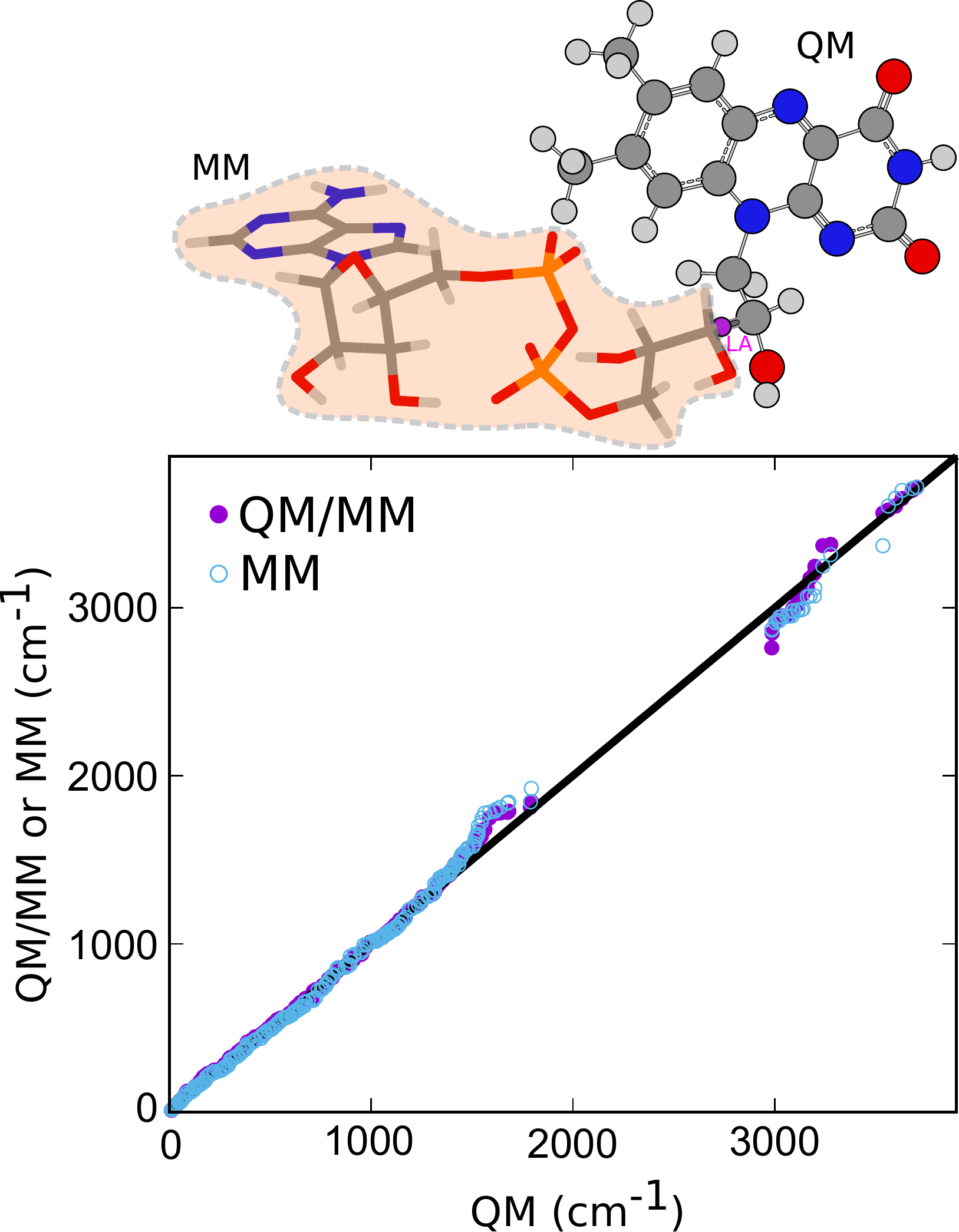}
    \caption{Comparison of the QM/MM and MM frequencies (y-axis) with respect to full QM frequencies (x-axis) of flavin adenine dinucleotide. For the computation of frequencies, a geometry optimization has been performed at each level of theory (QM: B3LYP/6-31G*, MM: Amber99; QM/MM: B3LYP/6-31G*/Amber99). For the QM/MM calculation, a link atom has been placed between the isoalloxazine ring (treated at the QM level) and the rest of the molecule as indicated in the graph.}
    \label{fig1}
\end{figure}

As a first example of our method, we have computed the frequencies of flavin adenine dinucleotide at the QM/MM level. In the QM region, we have placed the isoalloxazine ring while the rest of the molecule is considered MM. This selection requires the introduction of a link atom to separate the QM and MM regions, which has been placed 3 bonds far away from the isoalloxazine ring to avoid overpolarization, as shown in Fig.\ \ref{fig1}. In this figure, we show the frequencies computed at the QM/MM level compared to the QM frequencies, which we take as reference. For the sake of comparison, we also show the pure MM frequencies. Overall, the quality of the QM/MM and MM frequencies is in very good accordance with the QM frequencies. This is expected since the MM parameters have been obtained from purely \textit{ab initio} data. Some appreciable errors are observed for the highest frequency modes, as well as the mid-range. The QM/MM frequencies follow a similar trend than the MM frequencies, indicating that these modes have simply a small mixing and remain rather purely MM or purely QM. Indeed, the most delocalized frequencies are found only in the low frequency range (see further below). This results indicate that the choice of force field is indeed critical for obtaining a good spectrum of frequencies that is comparable to the \textit{ab initio} spectrum.

\begin{figure}
    \includegraphics[scale=0.8,keepaspectratio]{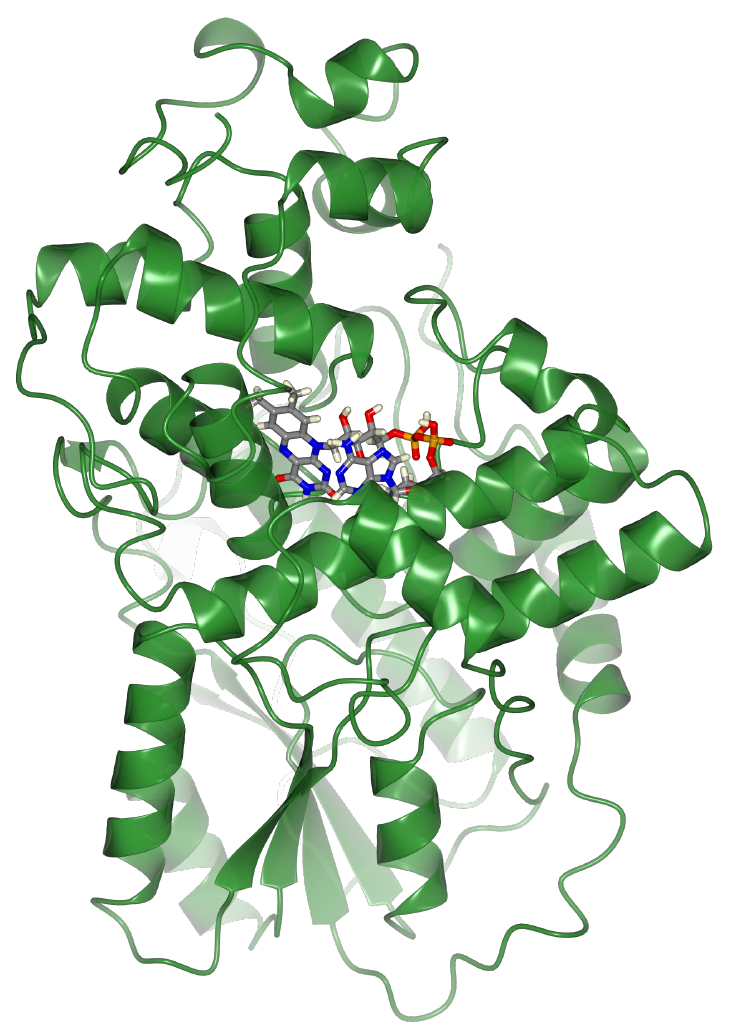}
    \caption{Representation of the \textit{arabidopsis thaliana} cryptochrome (PDBID: 2J4D) used in the QM/MM simulations. The FAD fragment (ball and stick) constitutes the QM subsystem, while the protein (ribbon) is treated at the MM level. Crystallographic water molecules have been excluded from the plot, but included in the simulations.}
    \label{fig2}
\end{figure}

To further illustrate the potential of our method, we have performed a normal mode analysis of plant cryptochrome, including vibronic coupling and homogeneous broadening of absorption spectrum using vibrationally resolved spectra simulations. For this, we have selected chain A of \textit{arabidopsis thaliana} cryptochrome (PDB code 2J4D, see Fig. \ref{fig2}). Crystallographic water molecules have been kept in the model. Hydrogens have been added for standard aminoacid protonation states at physiological pH. The entire FAD coenzyme was defined as the QM subsystem. FAD is not covalently linked to the rest of the protein, thus not requiring any special care of the QM/MM boundary. The initial PDB structure was pre-optimized at the Amber99 level using microiterations. Subsequent geometry optimization, hessian calculation and normal mode analysis were carried out at B3LYP/TZVP/TZVPFit/Amber99 level for the entire protein model.

One of the main advantages of the method presented here is that the CPSCF equations for the MM perturbations are not scaling with the MM subsystem size. As we showed, we can decouple this dependence by solving the Q-vector equations for N$_{QM}$ atoms and reconstruct the atomic charge derivatives with respect to MM perturbations by contracting with the external field derivatives. This induces an obvious saving in computational time. For the present case (plant cryptochrome, 9432 atoms at the B3LYP/TZVP/TZVPFit/Amber99 level using 32 cores of 2.6 GHz), the solution of conventional CPSCF equations for the full system would take around 86.5 hours, while the Q-vector method (84 atom) just takes 17 minutes. This allows the construction of the full Hessian including the charge polarization terms at the cost of a mechanical embedding QM/MM full Hessian.

\subsection{Normal mode analysis}

\begin{figure}
    \includegraphics[width=\linewidth,keepaspectratio]{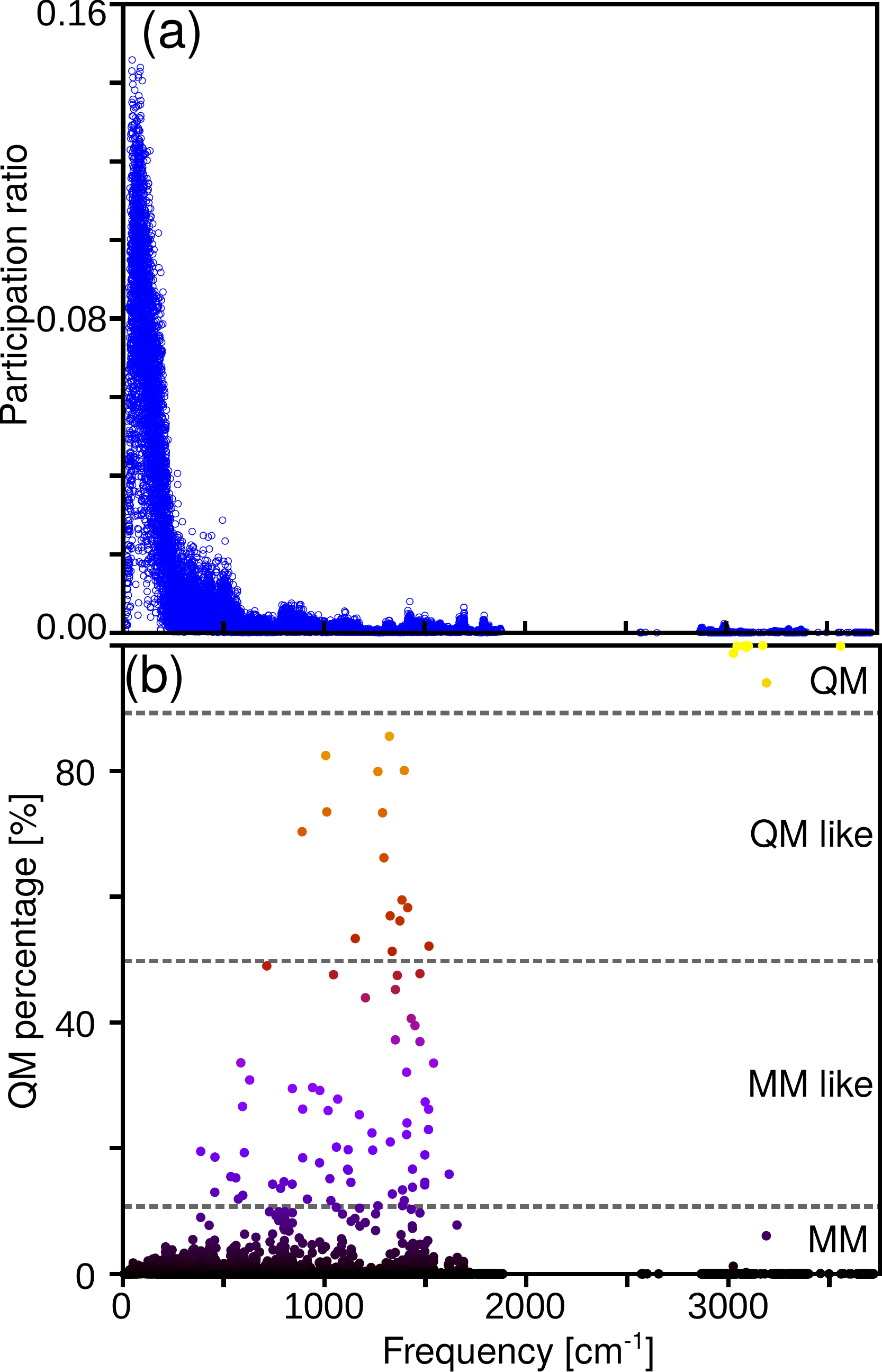}
    \caption{Normal mode analysis of plant cryptochrome. (a) participation ratio showing the degree of delocalization of normal modes, and (b) analysis of the character QM character. Modes with less than 10\% QM atom displacement are classified as MM (black) while modes with more than 90\% are classified as QM (yellow).}
    \label{fig3}
\end{figure}

We performed the normal mode analysis of the plant cryptochrome shown in Fig. \ref{fig2} at the minimum energy structure of the QM/MM internal energy. This geometry was obtained from a quadratically convergent optimization of the QM subsystem, with a steepest descent optimization of the MM susbsystem at every step of optimization (microiterations). At the minimum structure, the FAD remains in a U-folded conformation due to the electrostatic interactions with the protein environment. The protein model has 8121 atoms, from which 84 atoms are in the QM subsystem. This leads to a total of 24357 normal modes. In Fig. \ref{fig3}, we analyse the degree of delocalization as well as the QM character of these modes. First, we computed the participation ratio in order to measure the degree of delocalization. For this, we defined the participation ratio as
\begin{equation}
    PR(\omega_i)=\left(3N\cdot\sum_j^{3N}{(\vphantom{a}^0\mathrm{L}^\dagger_{i,j}\cdot \vphantom{a}^0\mathrm{L}_{j,i})^2}\right)^{-1}
\end{equation}
in which $\vphantom{a}^0{\mathrm{\bf L}}_i$ is the mass-weighted normal mode displacement of the ground state. For a fully delocalized mode, all atoms contribute equally to the mass-weighted displacement, and the participation ratio is equal to one, whereas if only few atoms contribute to the displacement, the participation ratio has the value of $(3N)^{-1}$ (which is close to 0 in the case of plant cryptochrome.) Following this analysis, we can observe that the low frequency modes (<200 cm$^{-1}$) show on average the largest degree of delocalization, involving thus a large number of atoms in each normal mode frequency. These modes are usually involved in describing protein phase transitions. The rest of the normal modes involve only a few atoms.

The participation ratio analysis is not sufficient to determine the degree of mixing between QM and MM atom motions. If one would perform a QM subsystem normal mode analysis, one would obtain 246 full QM normal modes. For the full protein diagonalization, these modes are strongly mixed. In Fig. \ref{fig3}, we report the QM character of each normal mode, defined as
\begin{equation}
    QM(\omega_i)=\sum_{j\in QM}^{3N}{\left(\vphantom{a}^0\mathrm{L}^\dagger_{i,j}\right)^2} .
\end{equation}
As we can observe, the low frequency modes below 200 cm$^{-1}$ have a negligible QM character. This implies a strong mixing between the QM and MM atoms in this region. Since the number of MM atoms is so large, the QM character is ``diluted'' within the MM motion. Only two normal modes  between 1200-1400 cm$^{-1}$ keep a clear QM character (>90\%), and another 23 normal modes have a QM character larger than 60\%, comprised in the range of 400-1600 cm$^{-1}$. Therefore, only 10\% of the FAD normal modes keep in essence its original character. For the rest, QM and MM atom motions are essentially mixed, even if the modes are not fully delocalized over the whole protein. Importantly, this mixing is present in the whole range of frequencies, and it is not only appearing for the fully delocalized low-frequency modes.

\subsection{Vibronic couplings, Huang-Rhys factors and spectral densities}

The Huang-Rhys couplings are a fundamental measure for the electron-nuclear (vibronic) coupling. The HR factor is directly related to the reorganization energy of one mode in the excited state. They can be defined as
\begin{equation}
\mathrm{HR}_{i,I}=\frac{1}{2}\omega_i\Delta q_{i,I}^2 \, ,    
\end{equation}
in which $\omega_i$ is the normal mode frequency and $\Delta \vphantom{a}q_{i,I}=\vphantom{a}^0{\mathrm{\bf L}}_{i}{\bf M}^{1/2}\left({\bf R}_I-{\bf R}_0\right)$ is the projection in the ground state normal mode direction $\vphantom{a}^0{\bf \mathrm{L}}_{i}$ of the Cartesian geometry difference between excited (${\bf R}_I$) and ground state (${\bf R}_0$) minima,  mass weighted by the mass metric ${\bf M}=\delta_{ij}m_i$, in which $m_i$ is the atomic mass of atom $i$. 

The $\Delta q^I_i$ can be approximately obtained by the so-called vertical gradient approximation, in which the mass-weighted energy gradient of the excited state I ($\Delta q^I_i$) is projected to the normal mode direction, leading to 
\begin{equation}
    \Delta q_{i,I}=-\frac{\left({\bm \nabla}{\bf E}_I|\vphantom{a}^0{\mathrm{\bf L}}_{i}\right)}{\omega_i^2}
    \label{eq:vgradient}
\end{equation}
In order to define the effect of the bath modes, the notion of spectral density $J(\omega)$ is useful. It can be defined as
\begin{equation}
    J_I(\omega)=\pi\sum_i{\mathrm{HR}_{i,I}{\omega_i^2}\delta(\omega-\omega_i)}
\end{equation}
Indeed, the spectral density contains information about each normal mode frequency, and its coupling to the system, which suffices to characterize the internal dissipation of an excited state due to nuclear motion. 

\begin{figure}
    \includegraphics[width=\linewidth,keepaspectratio]{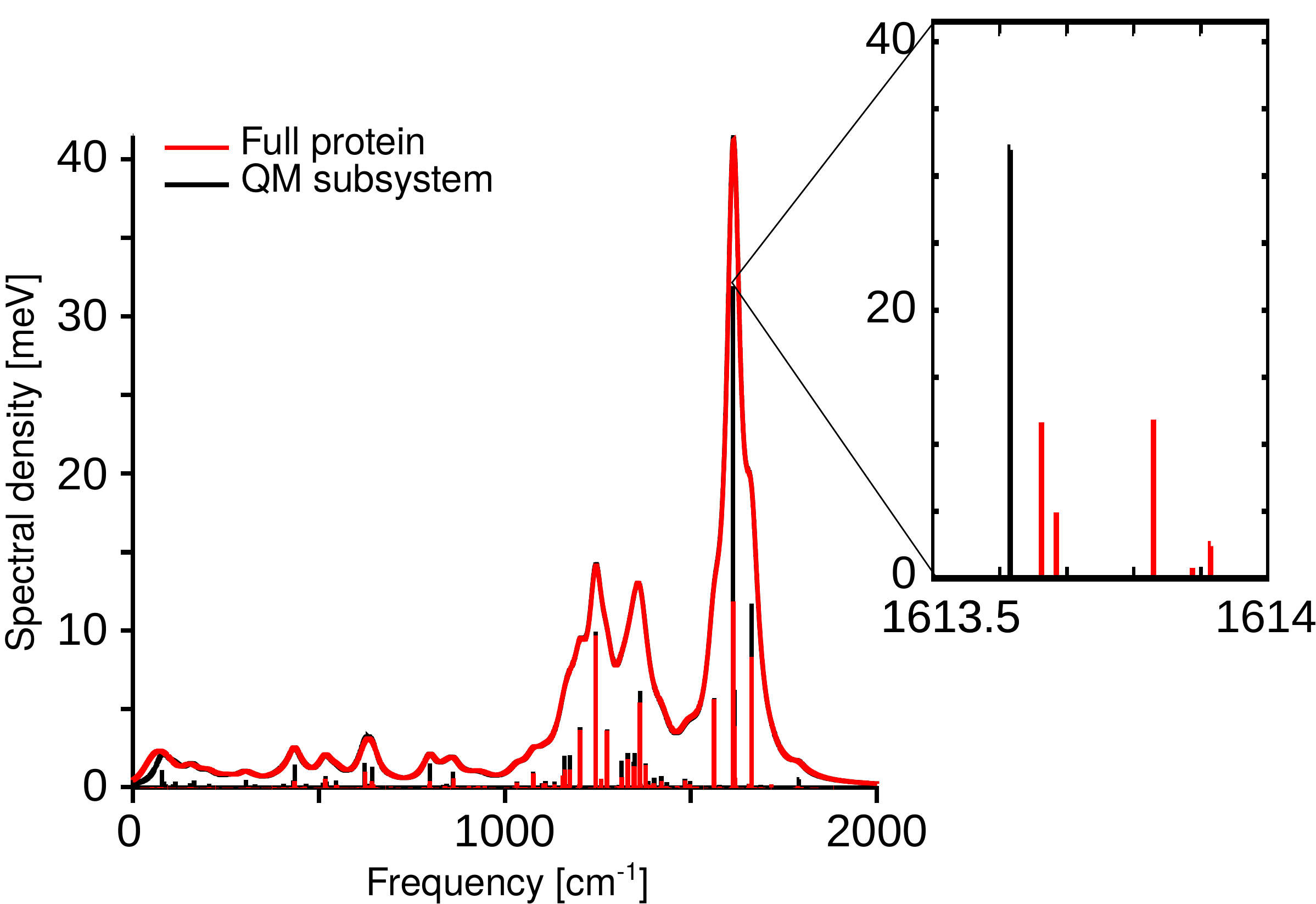}
    \caption{Stick spectrum corresponding to the spectral density for the first bright excited state. The continuous spectral density has been obtained from a Lorenzian broadening of 500 cm$^{-1}$. The red line correspond to the normal modes of the full protein, while the black line to the QM subsystem normal modes. The zoom-in of the most intense peak is shown in the right-hand inset.}
    \label{fig4}
\end{figure}

Using the information of the QM/MM analytic second derivatives together with the excited-state gradient at the ground-state minimum, we can construct the spectral densities characterizing the same cryptochrome model. In Figure \ref{fig4}, we have plotted such spectral density for the lowest energy bright excited state. Two spectral densities are shown, one corresponding to the normal modes and gradient of the full protein, and the other corresponding to the normal modes and gradient of the QM subsystem. 

The overall contribution of MM atoms to the broadened spectral density is only minor. This is expected because the transition density is mainly located on the chromophore, which is entirely located in the QM subsystem, and thus only modes with substantial QM motion should be contributing to the spectral density. Still, an indirect contribution of the MM atoms appear in the spectral density due to the mode mixing, as shown for example in the most intense peak around 1620-1625 cm$^{-1}$. While the QM spectral density has only one peak of 40 meV, the full protein spectral density shows a spectral density contribution of 4 modes. This is due to the mixing of QM and MM atom motions, alike to the intensity borrowing mechanism of electronic spectra, in which dark states couple to bright states borrowing dipole intensity and thus becoming observable. This can be observed for almost all peaks in the spectral density, which reflects once again the strong normal mode mixed character that we showed in Fig. \ref{fig3}. It is important to notice though that even though there is a strong mixing, the spectral densities of the QM and the full protein are essentially equivalent. This is because the normal mode mixing occurs only between QM and MM modes at almost the same frequency. Eventually, the effect of mixing has only a minor impact on the frequency of the modes (less than 1 cm$^{-1}$ for the most intense peak). This results justify the extraction of vibronic couplings from the normal mode analysis of the QM subsystem.

\subsection{Vibrationally resolved absorption spectrum}

Flavoproteins like cryptochrome act as blue-light sensors in biological systems.\cite{Conrad2014} These proteins have a well-resolved vibrational progression of the absorption spectrum, characteristic of the isoalloxazine chromophore. Unlike other types of proteins in which inhomogeneous broadening dominates, the homogeneous broadening has to be explicitly taken into account to reproduce the color of absorption of the flavin chromophore.\cite{Kabir2019,Mondal2020} The approximate vibronic couplings shown in Eq.\ \ref{eq:vgradient} can be used to compute the absorption spectrum including the vibrational structure due to homogeneous broadening. 

\begin{figure}
    \includegraphics[width=\linewidth,keepaspectratio]{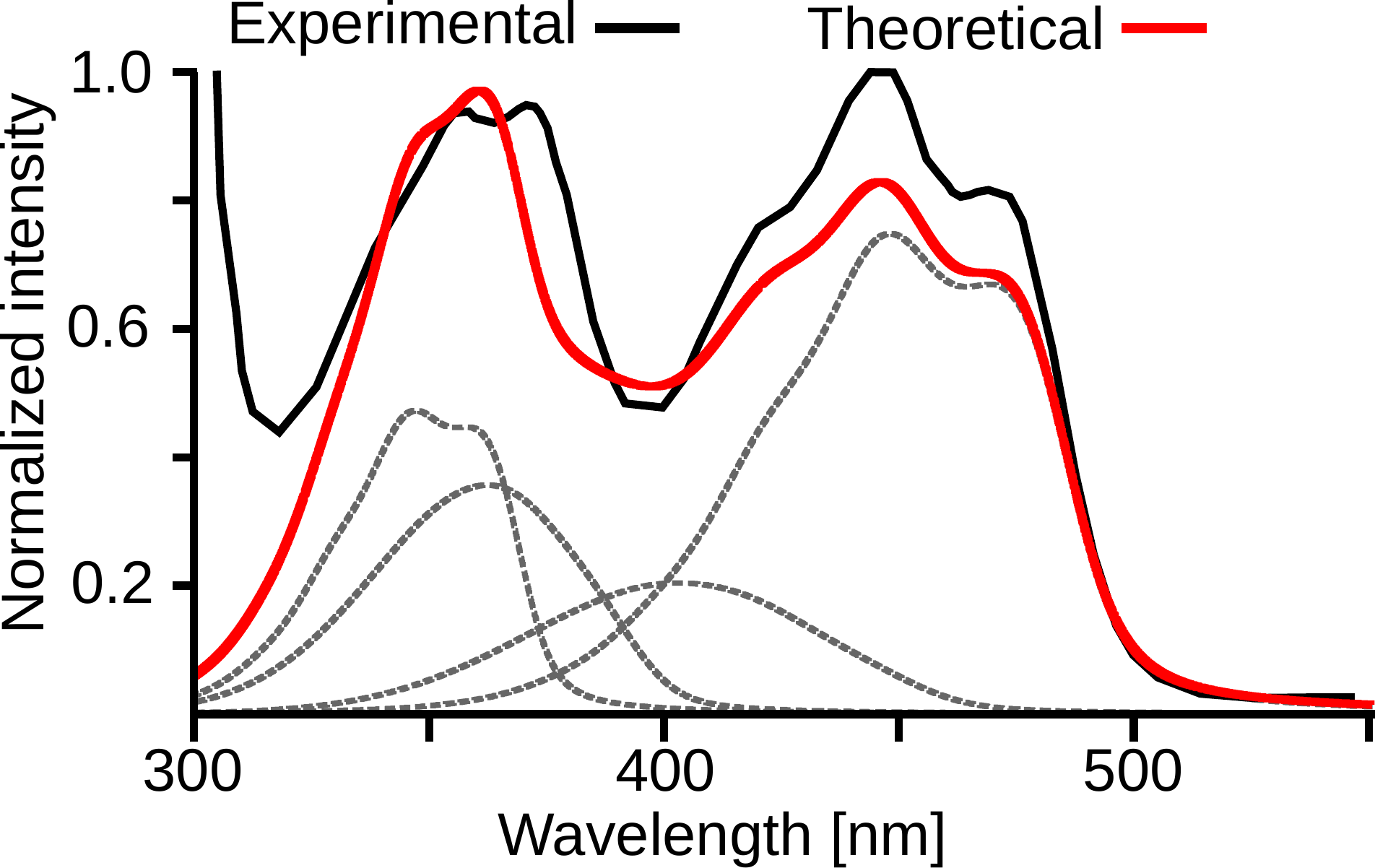}
    \caption{Comparison of the simulated vibrationally resolved absorption spectrum (red) with the experimental spectrum (black) of \textit{arabidopsis thaliana}. The grey dotted lines indicate the individual contributions from each electronic state. For the sake of comparison, both spectra have been normalized for the highest intensity peak between the range of 350 nm and 500 nm. The simulated spectrum has been red-shifted by 12 nm to match the experimental maximum. Experimental spectrum has been taken from Ref. \citenum{Muller14}}
    \label{fig5}
\end{figure}

Here, we compute the vibrational resolved spectrum of FAD in plant cryptochrome, by using the QM subsystem normal modes in conjunction with the vibrationally-resolved time-dependent approach of Bloino and coworkers.\cite{Baiardi13} In Fig.\ \ref{fig5}, we compare the experimental and theoretical absorption spectra for \textit{arabidopsis thaliana} plant cryptochrome with FAD in its fully oxidized form.\cite{Muller14} The experimental absorption spectrum is characterized by two main absorption peaks of $\pi\rightarrow\pi^*$ character. The lowest bright transition has a peak maximum around 450 nm and the second bright state at around 360 nm.\cite{Muller14} These peaks have a marked vibrational structure, especially due to in-plane C-N stretchings.\cite{Klaumunzer2010} Indeed, the first electronic transition is distributed in resolved peaks, approximately found at 470, 450 and 420 nm. Our simulated spectrum perfectly matches the relative position and intensities of the experimental peaks. From our simulations, we observe that these peaks are in fact formed from two electronic transitions, a HOMO$\rightarrow$LUMO transition which give the basic vibrational structure, and a HOMO-1$\rightarrow$LUMO transition which is centered at 420 nm. For the second absorption peak, we have a reasonable agreement between simulated and experimental spectra. Indeed, the relative peak position and the relative intensities between the two vibronic peaks are well reproduced by the theory. However, our theoretical simulations underestimate by around 40\% the relative intensity with respect to the lowest energy absorption peak. We attribute this either to the approximations of the level of theory or some missing contributions of low-intensity high-energy peak.

\section{Conclusion}

We have derived a fully analytic expression for the second derivative of the electrostatic embedding ESPF QM/MM energy. In this approach, the atomic charges representing the QM charge distribution are not only polarizable by the usual QM/MM electrostatic interactions, but also by the motion of the MM atoms. QM/MM electrostatic energy second derivative contains MM-block contributions too expensive to compute, and usually simply ignored or approximated. Using the ESPF method, we have shown that these terms can be equivalently computed with a set of auxiliary coupled-perturbed self-consistent field equations that only scales with the QM subsystem size,\cite{Schwinn19} reducing thus the total computational cost of the full electrostatic embedding QM/MM hessian to a similar cost than mechanical embedding QM/MM hessian.

To show the capabilities of our method, we have computed the full hessian of \textit{arabidopsis thaliana} plant cryptochrome. We have evidenced that most of the FAD chromophore (QM) vibrational modes are strongly mixed with the modes of the protein, and only a few high-frequency modes around 1200-1500 cm$^{-1}$ keep a full QM character. In addition, we have computed the vibronic couplings and spectral densities for the bright excited state, together with the vibrationally resolved spectra of FAD in the protein. These applications open up the use of QM/MM hessians construct vibronic models for photoactive proteins and QM systems in complex environments.

Several limitations are currently imposed in our current model for a practical use. First, one of the main limitations of our QM/MM method is the lack of periodic boundary conditions to include solvent effects. In the future, we plan to implement our method in an extension of the ESPF QM/MM method to include Ewald summations.\cite{Holden2019} Meantime, the only strategy is to freeze a state of the solvent, and relax the protein within a solvation shell. Since the QM equations do not scale with the MM subsystem size, the calculations considering a large number of solvent molecules should be tractable at a reasonable computational cost. Second, the current simulation of vibrationally resolved absorption spectrum requires a positive semi-definite Hessian matrix. This implies that any initial geometry must be optimized to a minimum energy structure so that all frequencies are positive. The geometry optimization step is time-consuming, hence severely limiting the usual efficiency associated with the inhomogenous broadening as it is frequently reported in the literature, i.e., by selection of snapshots from a MD simulation and performing the QM/MM analysis at each of them. This limitation could be overcome by using for example instantaneous normal modes,\cite{Kalstein2011} although further investigation in this direction is required.

Finally, we plan to extend our model to address computed vibrational spectroscopies. Such simulations face frequently the problem of comparing with experimental spectra, which could ultimately be attributed to the lack of anharmonic effects, both from the QM and the MM frequency spectrum. To better match the QM frequencies to experimental data, it is common to scale the Hessian via a scaling factor that depends on the method and the basis set. Morokuma and coworkers applied a similar strategy to their ONIOM QM/MM Hessian, in which three different scaling factors for the QM, MM and QM-MM Hessian blocks are applied.\cite{Dapprich99} Such scheme would be readily applied to our analytic QM/MM Hessian too. In addition, some recent developments at the level of force fields for better describing infra-red spectra can also be trivially integrated in our Hessian scheme.\cite{Nutt2003,Mackie2015}

\subsection{Appendix: Derivatives of the ESPF operator}
\label{sec:app1}

The ESPF atomic charge operator elements $Q_{A,\mu\nu}$ are obtained by fitting the corresponding electrostatic potential integrals $V_{k,\mu\nu}$,
\begin{equation}
V_{k,\mu\nu}=\int{d{\bf r}\chi_\mu({\bf r})|{\bf r}-{\bf R}_k|^{-1}\chi_\nu({\bf r})} 
\end{equation}
calculated over a grid of $N_{grid}$ points $k$, at position ${\bf R}_k$, around the QM atoms:
\begin{equation}
\label{eq:espfv}
\sum_{A}{T_{A,k}Q_{A,\mu\nu}}=V_{k,\mu\nu} 
\end{equation}
The kernel ${\bf T}$ is a rectangular matrix of dimensions $N_{grid}\cdot N_{QM}$. The kernel definition depends on the multipolar expansion order. For example, for atomic charges it contains the inverse distance between QM atom and the grid point $T_{A,k}=\left|{\bf R}_A-{\bf R}_k\right|^{-1}$. Thus, we can give an expression for the  $Q_{A,\mu\nu}=\langle\chi_\mu|\widehat{Q}_A|\chi_\nu\rangle$ atomic charge operator elements by solving the previous set of linear equations, leading to
\begin{equation}
Q_{A,\mu\nu}=\sum_{k}^{N_{grid}}{\left[\left({{\bf T}^\dagger{\bf T}}\right)^{-1}{\bf T}^\dagger\right]_{A,k} V_{k,\mu\nu}} 
\end{equation}

The derivative of the ESPF hamiltonian involves the derivative of the atomic charge operator. A recursive formula for the n-derivative of the charge operator can be obtained by directly deriving n-times Eq. \ref{eq:espfv}, 
\begin{equation}
Q_{A,\mu\nu}^{n}=\sum_{k}^{N_{grid}}{\left[\left({{\bf T}^\dagger{\bf T}}\right)^{-1}{\bf T}^\dagger\right]_{A,k}{F}_{k,\mu\nu}^n} 
\end{equation}
in which 
\begin{equation}
    {F}_{k,\mu\nu}^n=V^{n}_{k,\mu\nu}-\sum_{i=0}^{n-1}{\sum_B^{N_{QM}}{T^{n-i}_{k,B}Q^i_{B,\mu\nu}}} 
\end{equation}

Here we give the explicit first and second derivatives of the ${\bf h}$ operator defined in Eq. \ref{eq:hespf}. The first derivative of the operator is given by
\begin{equation}
h_{\mu\nu}^{x}=\sum_A{\left(Q^{x}_{A,\mu\nu}\phi_A+Q_{A,\mu\nu}\phi_A^{x}\right)} 
\end{equation}
The second term involves the derivative of the classical external field, which have simple expression given the form of the force field. The derivative of the atomic charge operator can be in term expanded (in matrix notation) as
\begin{equation}
{\bf Q}^{x}=\left({{\bf T}^\dagger{\bf T}}\right)^{-1}{\bf T}^\dagger\left[{\bf V}^x-{\bf T}^x{\bf Q}\right]
\end{equation}
we can easily see that the computation of first-order derivatives of ${\bf h}$ operators requires the computation of $\phi_A^{x}$, $V^x_{k,\mu\nu}$ and ${\bf T}^{x}$ in addition to the zero-order field and charges.

The second derivatives of the ${\bf h}$ operator is given by
\begin{eqnarray}
h_{\mu\nu}^{xy}&=&\sum_A{\left(Q^{xy}_{A,\mu\nu}\phi_A+Q_{A,\mu\nu}\phi_A^{xy}\right)} \\
&+&\sum_A{\left(Q^{x}_{A,\mu\nu}\phi_A^{y}+Q^{y}_{A,\mu\nu}\phi_A^{x}\right)}  \nonumber
\end{eqnarray}
The second term involves the second derivative of the classical external field, which again has a simple expression given the form of the force field, while the third and fourth term are constructed with elements of the first derivative of the operator. Hence, one needs only to develop the second derivative of the atomic charge operator. This is given by
\begin{eqnarray}
{\bf Q}^{xy}&=&\left({{\bf T}^\dagger{\bf T}}\right)^{-1}{\bf T}^\dagger\left[{\bf V}^{xy}-{\bf T}^{xy}{\bf Q}\right]\nonumber \\
&-&\left({{\bf T}^\dagger{\bf T}}\right)^{-1}{\bf T}^\dagger\left[{\bf T}^x{\bf Q}^y+{\bf T}^y{\bf Q}^x\right]\nonumber \\
\end{eqnarray}
Therefore, in order to obtain the second derivative of the ESPF operator, we need to compute $\phi_A^{xy}$, $V^{xy}_{k,\mu\nu}$ and ${\bf T}^{xy}$ in addition to the first derivatives and the zero-order quantities.

\begin{acknowledgement}
The authors acknowledge financial support by the ``Agence  Nationale  pour  la  Recherche'' through the project BIOMAGNET (ANR-16CE29-0008-01). ``Centre de Calcul Intensif d'Aix-Marseille'' is acknowledged for granting access to its high performance computing resources.
\end{acknowledgement}

\bibliography{refs}

\end{document}